\documentclass[reqno,12pt,a4paper]{amsart}

\usepackage{mathrsfs}
\usepackage{amssymb}
\usepackage{amsfonts}
\usepackage{latexsym}
\usepackage{amsthm}
\usepackage{xcolor}
\usepackage{wrapfig}
\usepackage{floatflt}

\usepackage{graphicx}
\def\lb{\label}

\newcommand{\er}[1]{\textrm{(\ref{#1})}}

\begin{document}


\renewcommand{\theequation}{\arabic{section}.\arabic{equation}}
\theoremstyle{plain}
\newtheorem{theorem}{\bf Theorem}[section]
\newtheorem{lemma}[theorem]{\bf Lemma}
\newtheorem{corollary}[theorem]{\bf Corollary}
\newtheorem{proposition}[theorem]{\bf Proposition}
\newtheorem{definition}[theorem]{\bf Definition}
\newtheorem{remark}[theorem]{\it Remark}

\def\a{\alpha}  \def\cA{{\mathcal A}}     \def\bA{{\bf A}}  \def\mA{{\mathscr A}}
\def\b{\beta}   \def\cB{{\mathcal B}}     \def\bB{{\bf B}}  \def\mB{{\mathscr B}}
\def\g{\gamma}  \def\cC{{\mathcal C}}     \def\bC{{\bf C}}  \def\mC{{\mathscr C}}
\def\G{\Gamma}  \def\cD{{\mathcal D}}     \def\bD{{\bf D}}  \def\mD{{\mathscr D}}
\def\d{\delta}  \def\cE{{\mathcal E}}     \def\bE{{\bf E}}  \def\mE{{\mathscr E}}
\def\D{\Delta}  \def\cF{{\mathcal F}}     \def\bF{{\bf F}}  \def\mF{{\mathscr F}}
\def\c{\chi}    \def\cG{{\mathcal G}}     \def\bG{{\bf G}}  \def\mG{{\mathscr G}}
\def\z{\zeta}   \def\cH{{\mathcal H}}     \def\bH{{\bf H}}  \def\mH{{\mathscr H}}
\def\e{\eta}    \def\cI{{\mathcal I}}     \def\bI{{\bf I}}  \def\mI{{\mathscr I}}
\def\p{\psi}    \def\cJ{{\mathcal J}}     \def\bJ{{\bf J}}  \def\mJ{{\mathscr J}}
\def\vT{\Theta} \def\cK{{\mathcal K}}     \def\bK{{\bf K}}  \def\mK{{\mathscr K}}
\def\k{\kappa}  \def\cL{{\mathcal L}}     \def\bL{{\bf L}}  \def\mL{{\mathscr L}}
\def\l{\lambda} \def\cM{{\mathcal M}}     \def\bM{{\bf M}}  \def\mM{{\mathscr M}}
\def\L{\Lambda} \def\cN{{\mathcal N}}     \def\bN{{\bf N}}  \def\mN{{\mathscr N}}
\def\m{\mu}     \def\cO{{\mathcal O}}     \def\bO{{\bf O}}  \def\mO{{\mathscr O}}
\def\n{\nu}     \def\cP{{\mathcal P}}     \def\bP{{\bf P}}  \def\mP{{\mathscr P}}
\def\r{\rho}    \def\cQ{{\mathcal Q}}     \def\bQ{{\bf Q}}  \def\mQ{{\mathscr Q}}
\def\s{\sigma}  \def\cR{{\mathcal R}}     \def\bR{{\bf R}}  \def\mR{{\mathscr R}}
\def\S{\Sigma}  \def\cS{{\mathcal S}}     \def\bS{{\bf S}}  \def\mS{{\mathscr S}}
\def\t{\tau}    \def\cT{{\mathcal T}}     \def\bT{{\bf T}}  \def\mT{{\mathscr T}}
\def\f{\phi}    \def\cU{{\mathcal U}}     \def\bU{{\bf U}}  \def\mU{{\mathscr U}}
\def\F{\Phi}    \def\cV{{\mathcal V}}     \def\bV{{\bf V}}  \def\mV{{\mathscr V}}
\def\P{\Psi}    \def\cW{{\mathcal W}}     \def\bW{{\bf W}}  \def\mW{{\mathscr W}}
\def\o{\omega}  \def\cX{{\mathcal X}}     \def\bX{{\bf X}}  \def\mX{{\mathscr X}}
\def\x{\xi}     \def\cY{{\mathcal Y}}     \def\bY{{\bf Y}}  \def\mY{{\mathscr Y}}
\def\X{\Xi}     \def\cZ{{\mathcal Z}}     \def\bZ{{\bf Z}}  \def\mZ{{\mathscr Z}}
\def\O{\Omega}

\newcommand{\gA}{\mathfrak{A}}
\newcommand{\gB}{\mathfrak{B}}
\newcommand{\gC}{\mathfrak{C}}
\newcommand{\gD}{\mathfrak{D}}
\newcommand{\gE}{\mathfrak{E}}
\newcommand{\gF}{\mathfrak{F}}
\newcommand{\gG}{\mathfrak{G}}
\newcommand{\gH}{\mathfrak{H}}
\newcommand{\gI}{\mathfrak{I}}
\newcommand{\gJ}{\mathfrak{J}}
\newcommand{\gK}{\mathfrak{K}}
\newcommand{\gL}{\mathfrak{L}}
\newcommand{\gM}{\mathfrak{M}}
\newcommand{\gN}{\mathfrak{N}}
\newcommand{\gO}{\mathfrak{O}}
\newcommand{\gP}{\mathfrak{P}}
\newcommand{\gQ}{\mathfrak{Q}}
\newcommand{\gR}{\mathfrak{R}}
\newcommand{\gS}{\mathfrak{S}}
\newcommand{\gT}{\mathfrak{T}}
\newcommand{\gU}{\mathfrak{U}}
\newcommand{\gV}{\mathfrak{V}}
\newcommand{\gW}{\mathfrak{W}}
\newcommand{\gX}{\mathfrak{X}}
\newcommand{\gY}{\mathfrak{Y}}
\newcommand{\gZ}{\mathfrak{Z}}

\newcommand{\ba}{\mbox{\boldmath$\alpha$}}
\newcommand{\bk}{\mbox{\boldmath$\kappa$}}
\newcommand{\bm}{\mbox{\boldmath$\mu$}}
\newcommand{\bet}{\mbox{\boldmath$\eta$}}

\def\ve{\varepsilon}   \def\vt{\vartheta}    \def\vp{\varphi}    \def\vk{\varkappa}

\def\Z{{\mathbb Z}}    \def\R{{\mathbb R}}   \def\C{{\mathbb C}}    \def\K{{\mathbb K}}
\def\T{{\mathbb T}}    \def\N{{\mathbb N}}   \def\dD{{\mathbb D}}


\def\la{\leftarrow}              \def\ra{\rightarrow}            \def\Ra{\Rightarrow}
\def\ua{\uparrow}                \def\da{\downarrow}
\def\lra{\leftrightarrow}        \def\Lra{\Leftrightarrow}


\def\lt{\biggl}                  \def\rt{\biggr}
\def\ol{\overline}               \def\wt{\widetilde}
\def\no{\noindent}


\let\ge\geqslant                 \let\le\leqslant
\def\lan{\langle}                \def\ran{\rangle}
\def\/{\over}                    \def\iy{\infty}
\def\sm{\setminus}               \def\es{\emptyset}
\def\ss{\subset}                 \def\ts{\times}
\def\pa{\partial}                \def\os{\oplus}
\def\om{\ominus}                 \def\ev{\equiv}
\def\iint{\int\!\!\!\int}        \def\iintt{\mathop{\int\!\!\int\!\!\dots\!\!\int}\limits}
\def\el2{\ell^{\,2}}             \def\1{1\!\!1}
\def\sh{\sharp}
\def\wh{\widehat}
\def\bs{\backslash}

\def\all{\mathop{\mathrm{all}}\nolimits}
\def\Area{\mathop{\mathrm{Area}}\nolimits}
\def\arg{\mathop{\mathrm{arg}}\nolimits}
\def\const{\mathop{\mathrm{const}}\nolimits}
\def\det{\mathop{\mathrm{det}}\nolimits}
\def\diag{\mathop{\mathrm{diag}}\nolimits}
\def\diam{\mathop{\mathrm{diam}}\nolimits}
\def\dim{\mathop{\mathrm{dim}}\nolimits}
\def\dist{\mathop{\mathrm{dist}}\nolimits}
\def\Im{\mathop{\mathrm{Im}}\nolimits}
\def\Iso{\mathop{\mathrm{Iso}}\nolimits}
\def\Ker{\mathop{\mathrm{Ker}}\nolimits}
\def\Lip{\mathop{\mathrm{Lip}}\nolimits}
\def\rank{\mathop{\mathrm{rank}}\limits}
\def\Ran{\mathop{\mathrm{Ran}}\nolimits}
\def\Re{\mathop{\mathrm{Re}}\nolimits}
\def\Res{\mathop{\mathrm{Res}}\nolimits}
\def\res{\mathop{\mathrm{res}}\limits}
\def\sign{\mathop{\mathrm{sign}}\nolimits}
\def\span{\mathop{\mathrm{span}}\nolimits}
\def\supp{\mathop{\mathrm{supp}}\nolimits}
\def\Tr{\mathop{\mathrm{Tr}}\nolimits}
\def\BBox{\hspace{1mm}\vrule height6pt width5.5pt depth0pt \hspace{6pt}}
\def\where{\mathop{\mathrm{where}}\nolimits}
\def\as{\mathop{\mathrm{as}}\nolimits}


\newcommand\nh[2]{\widehat{#1}\vphantom{#1}^{(#2)}}
\def\dia{\diamond}

\def\Oplus{\bigoplus\nolimits}



\def\qqq{\qquad}
\def\qq{\quad}
\let\ge\geqslant
\let\le\leqslant
\let\geq\geqslant
\let\leq\leqslant
\newcommand{\ca}{\begin{cases}}
\newcommand{\ac}{\end{cases}}
\newcommand{\ma}{\begin{pmatrix}}
\newcommand{\am}{\end{pmatrix}}
\renewcommand{\[}{\begin{equation}}
\renewcommand{\]}{\end{equation}}
\def\eq{\begin{equation}}
\def\qe{\end{equation}}
\def\[{\begin{equation}}
\def\bu{\bullet}

\title[{Periodic lattice with defects}]
        {Periodic lattice with defects}
\date{\today}

\def\Wr{\mathop{\rm Wr}\nolimits}
\def\BBox{\hspace{1mm}\vrule height6pt width5.5pt depth0pt \hspace{6pt}}

\def\Diag{\mathop{\rm Diag}\nolimits}

\date{\today}
\author[Anton Kutsenko]{Anton Kutsenko}
\address{Laboratoire de M\'ecanique Physique, UMR CNRS 5469,
Universit\'e Bordeaux 1, Talence 33405, France,  \qqq email \
aak@nxt.ru }

\begin{abstract}
The  discrete periodic lattice of masses and springs with line  and
point  defects is considered.  The  dispersion equations for
propagative, guided and  localised  waves  are obtained. The
detailed analysis of example with three masses is provided.
\end{abstract}

\maketitle

\section{Introduction}
\setcounter{equation}{0}

There are many papers devoted to the wave propagation through the
discrete periodic lattice with different types of defects. One of
the early works \cite{M} provides explicit solution of the wave
equation for the periodic lattice with two different masses. The
structure with an infinite line defect embedded in an infinite
square lattice has been considered in the paper \cite{OA}. In this
case the dispersion relations for localised modes can be computed in
explicit form. In the paper \cite{MS} the authors examined several
classes of continuous and discrete models with various defect
configurations. In the recent paper  \cite{CNJMM} the homogeneous
lattice perturbed by the finite line of masses is considered with
respect to analysis of eigenmodes. Also note the large number of
papers devoted to periodic structures without localised defects but
with the boundaries, see references in \cite{KK1}.

The motivation for the present paper is to combine perturbation of
the periodic lattice by a periodic line (like in \cite{OA}) with
that by a finite defect  (like in \cite{CNJMM}) and to obtain a new
dispersion equation for such configurations. The developed method
can be applied for the different types of lattices with arbitrary
horizontal and vertical periods. Note that this method is based on
the modified Green function approach which differs from the method
of \cite{M1}. Further we plan to adapt the so-called monodromy
matrix approach (see e.g. \cite{KS}) for obtaining dispersion
equations for localised modes. In particular the monodromy matrix
approach should allows us to obtain localised modes in propagative
and guided spectral intervals.

In the example below, we consider the homogeneous lattice with mass
of particles $M=1$ perturbed by the line defect with mass of
particles $\wt M$ and with one single mass $\ol{M}$. The detailed
analysis of the band-gap structure of propagative and guided spectra
and the necessary and sufficient conditions of existing of localised
modes are obtained. We note an interesting phenomenon: if the mass
$\wt M$ increases indefinitely then the upper limit of mass $\ol{M}$
for which the localised mode exists is exactly
$\frac34-\frac1{2\pi}$ (somewhat unusual appearance of $\pi$).

The work is organized as follows. Section \ref{S2} contains our main
results, namely, the dispersion equations for three types of
defects: 1) periodic lattice without defects 2) periodic lattice
with periodic strip 3) periodic lattice with periodic strip and with
finite defects. In Section \ref{S4} the derivations of dispersion
equations from the Section \ref{S2} are provided. In Section
\ref{S3} the propagative, guided and localised spectra for the
homogeneous lattice with line and single defect are analised.

\section{Dispersion equations}
\setcounter{equation}{0}

\lb{S2}

\begin{figure}[h]
\center{\includegraphics[width=0.9\linewidth]{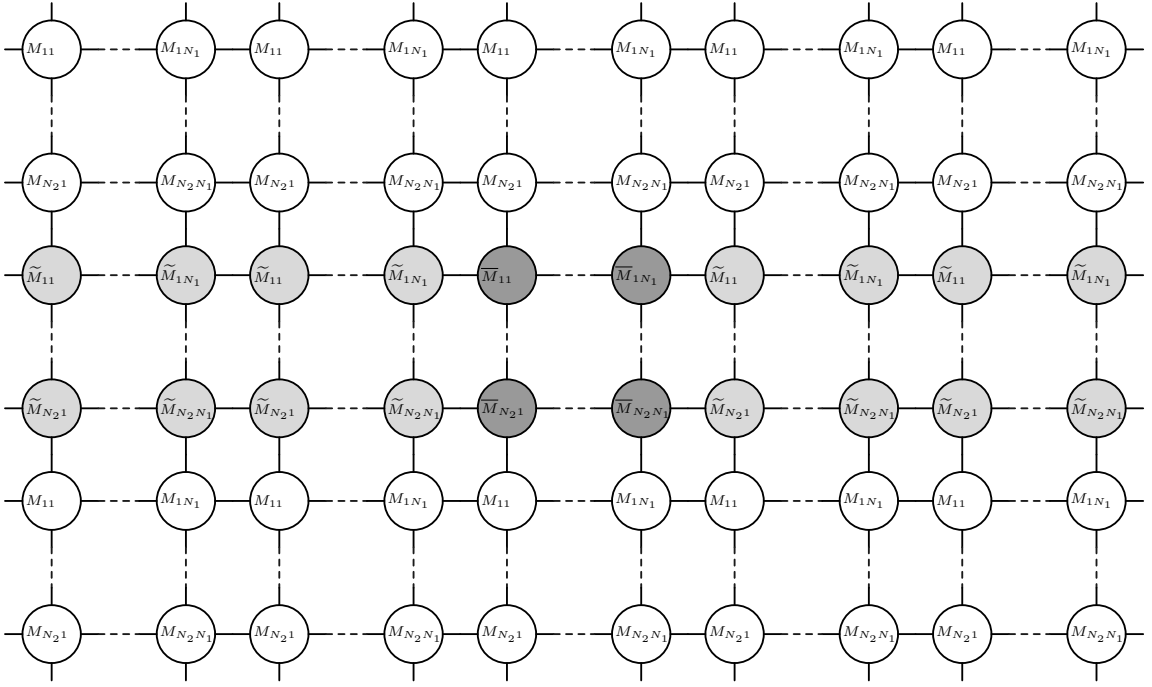}}
\caption{Periodic lattice with defects.} \label{fig0}
\end{figure}


\subsection{Periodic lattice.}
Consider the 2D periodic lattice (see Fig \ref{fig0}) without
defects, i.e. $M_{\bf n}=\wt M_{\bf n}=\ol M_{\bf n}$. The wave
equation  takes the following form
\[\lb{001}
 \sum_{{\bf n}'\sim{\bf n}}u_{{\bf n}'}-4u_{\bf n}=-\o^2M_{{\bf n}}u_{{\bf n}},
\]
where $\o$ is the frequency, $u_{\bf n}$ is antiplane displacement
and ${\bf n}=(n_1,n_2)$ is a position of the mass $M_{\bf n}$ on the
square lattice (number of the nod). We suppose that all quantities
in \er{001} are normalised (dimensionless). The notation ${\bf
n}\sim{\bf n}'$ means that there exists a link between nodes with
numbers ${\bf n}$ and ${\bf n}'$. Suppose that the function $M_{\bf
n}$ is periodic
\[\lb{002}
 M_{{\bf n}+N_1{\bf e}_1+N_2{\bf e}_2}=M_{{\bf n}},\ \ \forall {\bf
 n}\in\Z^2
\]
with basis vectors ${\bf e}_1=(1,0)$, ${\bf e}_2=(0,1)$ and periods
$N_1,N_2\ge1$. After applying Fourier transformation (see
\er{103}-\er{105}) the equation \er{001} takes the following form
\[\lb{003}
 \hat{\bf L}\hat{\bf u}=-\o^2\hat{\bf M}\hat{\bf u}
\]
with finite matrices defined on the unit cell
$\cN=[1..N_1]\ts[1..N_2]$:
\[\lb{004}
 \hat{\bf M}=(M_{\bf n}\d_{{\bf n}{\bf n}'})_{{\bf n},{\bf n}'\in\cN},\ \ \hat{\bf L}=\hat{\bf
 L}_0-4\hat{\bf I},
\]
\[\lb{005}
 \hat{\bf L}_0=(\hat\d_{{\bf n}\sim{\bf n}'})_{{\bf n},{\bf
 n}'\in\cN},\ \ \hat{\bf I}=(\d_{{\bf n}{\bf n}'})_{{\bf n},{\bf
 n}'\in\cN},
\]
where $\d_{{\bf n}{\bf n}'}$ is a Kronecker symbol and
\[\lb{006}
 \hat\d_{{\bf n}\sim{\bf n}'}=\d_{{\bf n}\sim{\bf n}'}+
\]
$$
  e^{- ik_1}\d_{{\bf n}-N_1{\bf e}_1\sim{\bf n}'}+
  e^{ ik_1}\d_{{\bf n}+N_1{\bf e}_1\sim{\bf n}'}+
$$
$$
  e^{- ik_2}\d_{{\bf n}-N_2{\bf e}_2\sim{\bf n}'}+
  e^{ ik_2}\d_{{\bf n}+N_2{\bf e}_2\sim{\bf n}'}
$$
with $\d_{{\bf n}\sim{\bf n}'}=1$ iff there exists a link between
nodes ${\bf n}$ and ${\bf n}'$ and $\d_{{\bf n}\sim{\bf n}'}=0$
otherwise.

The {\it dispersion equation} which determines the Floquet branches
$\o_{\rm p}({\bf k})$ (dispersion curves for propagative spectrum)
can be written in the form
\[\lb{009}
 \det\hat{\bf L}_{\rm p}(\o,{\bf k})=0,\ \ {\rm where}\ \
 \hat{\bf L}_{\rm p}\ev\hat{\bf L}+\o^2\hat{\bf M}.
\]

\subsection{Periodic lattice with periodic strip.} Here the
periodic strip with masses $\wt M_{\bf n}=M_{\bf n}+M^{(1)}_{\bf n}$
(see Fig. \ref{fig0}) is added to our lattice. In this case along
with the propagative spectrum $\o_{\rm p}$ there appears the guided
spectrum $\o_{\rm g}$. The guided spectrum corresponds to the waves
which are bounded (quasiperiodic) along the added strip and decay
along  the direction which is perpendicular to the strip. The guided
dispersion curves depend on Floquet parameter $k_1$ and can be
determined from the {\it dispersion equation} (see the derivation in
Section \ref{S42})
\[\lb{010}
 \det\hat{\bf L}_{\rm g}(\o,k_1)=0{\rm\ \ where\ \ }
 \hat{\bf L}_{\rm g}\ev\hat{\bf I}+\o^2\langle\hat{\bf L}_{\rm p}^{-1}\rangle_2\hat{\bf M}_1
\]
and
\[\lb{011}
 \hat{\bf M}_1=(M_{{\bf n}}^{(1)}\d_{{\bf n}{\bf n}'})_{{\bf n},{\bf
 n}'\in\cN},\ \ \ \langle\cdot\rangle_j=\frac1{2\pi}\int_{-\pi}^{\pi}\cdot
 dk_j.
\]
Note that the equation \er{010} is not valid in the bands of
propagative spectrum, i.e. in the sets
\[\lb{012}
 I_{\rm p}(k_1)=\o_{\rm p}(k_1,[-\pi,\pi])
\]
which correspond to the projection of the propagative dispersion
curves $\o_{\rm p}({\bf k})$ on the plane $(\o,k_1)$. We need this
restriction because the inverse of ${\bf L}_{\rm p}$ (see \er{010})
does not exist in the intervals $I_{\rm p}(k_1)$.

\subsection{Periodic lattice with periodic strip and with localised
inclusions.} Here the localised inclusions with masses $\ol{M}_{\bf
n}=\wt M_{\bf n}+M^{(2)}_{\bf n}$ (see Fig. \ref{fig0}) are added to
our periodic lattice with the strip. In this case along with
propagative $\o_{\rm p}$ and guided $\o_{\rm g}$ spectra the
localised spectrum $\o_{\rm loc}$ appears. The localised spectrum
corresponds to the waves which decay in any direction of our
lattice. The localised spectrum $\o_{\rm loc}$ can be determined
from the following {\it dispersion equation} (see the derivation in
Section \ref{S42})
\[\lb{013}
 D_{\rm loc}(\o)\ev\det(\hat{\bf I}+\o^2\langle\hat{\bf L}_{\rm g}^{-1}\langle{\bf L}_{\rm
 p}^{-1}\rangle_2\rangle_1
 \hat{\bf
 M}_2)=0,
\]
where
\[\lb{014}
 \hat{\bf M}_2=(M_{{\bf n}}^{(2)}\d_{{\bf n}{\bf n}'})_{{\bf n},{\bf
 n}'\in\cN}.
\]
Note that the equation \er{013} is not valid in the bands of
propagative and guided spectra, i.e. in the sets
\[\lb{015}
 I_{\rm p}=I_{\rm p}([-\pi,\pi]),\ \ I_{\rm g}=\o_{\rm g}([-\pi,\pi])
\]
which correspond to the projection of propagative $\o_{\rm p}({\bf
k})$ and guided $\o_{\rm g}(k_1)$ dispersion curves on the axis
$\o$. We need this restriction because the inverse of ${\bf L}_{\rm
p}$ and ${\bf L}_{\rm g}$ (see \er{013}) do not exist in the
intervals $I_{\rm p}$ and $I_{\rm g}$.

\section{The derivation of dispersion equations}
\setcounter{equation}{0}

\lb{S4}

\subsection{Periodic lattice.}\lb{S41} The set of linear equations \er{001} can be represented in the form of
infinite matrices
\[\lb{100}
 {\bf L}{\bf u}=-\o^2{\bf M}{\bf u},\ \ {\rm where}\ \ {\bf u}=(u_{\bf
 n})_{{\bf n}\in\Z^2}
\]
and
\[\lb{101}
 {\bf M}=(M_{\bf n}\d_{{\bf n}{\bf n}'})_{{\bf n},{\bf n}'\in\Z^2},\ \ {\bf L}={\bf L}_0-4{\bf
 I},
\]
\[\lb{102}
 {\bf L}_0=(\d_{{\bf n}\sim{\bf n}'})_{{\bf n},{\bf n}'\in\Z^2},\ \
 {\bf I}=(\d_{{\bf n}{\bf n}'})_{{\bf n},{\bf
 n}'\in\Z^2}.
\]
The infinite system \er{100} can be rewritten in the Fourier space
as a finite system. For this we introduce the operator
\[\lb{103}
 \cF:\ell^2(\Z^2)\to L^2_{\cN}([-\pi,\pi]^2),
\ \
 \cF({\bf u})=\hat{\bf u}({\bf k}),
\]
where
\[\lb{104}
 \hat{\bf u}({\bf k})=(\hat u_{\bf n}({\bf k}))_{{\bf n}\in\cN},\ \
 \hat u_{\bf n}({\bf k})=
 \sum_{{\bf r}\in\Z^2}u_{{\bf n}+r_1N_1{\bf e}_1+r_2N_2{\bf e}_2}\exp(i{\bf r}\cdot{\bf
 k}).
\]
Under the action of the operator $\cF$ any sequence ${\bf u}$ from
$\ell^2(\Z^2)$ becomes the vector-function $\hat{\bf u}({\bf k})$
with $N_1N_2$ components, any component is a function from
$L^2([-\pi,\pi]^2)$. Then the wave equation \er{100} can be
rewritten in the form
\[\lb{105}
 \cF{\bf L}\cF^{-1}\hat{\bf u}=-\o^2\cF{\bf M}\cF^{-1}\hat{\bf u}
\]
which coincide with \er{003} because $\cF{\bf L}\cF^{-1}=\hat{\bf
L}$ and $\cF{\bf M}\cF^{-1}=\hat{\bf M}$.

\subsection{Periodic lattice with periodic strip.} \lb{S42}
In this case the wave equation \er{003} becomes
\[\lb{106}
 \hat{\bf L}\hat{\bf u}=-\o^2(\hat{\bf M}\hat{\bf u}+\hat{\bf M}_1\langle\hat{\bf
 u}\rangle_2)
\]
with $\hat{\bf M}_1$ defined in \er{011}. From the equation \er{106}
we obtain that
\[\lb{107}
 \hat{\bf u}=-\o^2\hat{\bf L}_{\rm p}^{-1}\hat{\bf M}_1\langle\hat{\bf
 u}\rangle_2
\]
with $\hat{\bf L}_{\rm p}$ defined in \er{009}. Integrating with
respect to $k_2$ we get
\[\lb{108}
 \langle\hat{\bf
 u}\rangle_2=-\o^2\langle\hat{\bf L}_{\rm p}^{-1}\rangle_2\hat{\bf M}_1\langle\hat{\bf
 u}\rangle_2
\]
or (see definition of $\hat{\bf L}_{\rm g}$ in \er{010})
\[\lb{109}
 \hat{\bf L}_{\rm g}\langle\hat{\bf
 u}\rangle_2=0.
\]
From \er{109} we obtain the condition of existence of guided waves
\er{010}.

For our purpose we need the following fact. Consider the equation
\[\lb{110}
 \hat{\bf L}\hat{\bf u}+\o^2\hat{\bf M}\hat{\bf u}+\o^2\hat{\bf
 M}_1\langle\hat{\bf u}\rangle_2=\hat{\bf f}
\]
with some vector-function $\hat{\bf f}$ from
$L^2_{\cN}([-\pi,\pi]^2)$. Then the solution of this equation takes
the following form
\[\lb{111}
 \hat{\bf u}={\bf L}_{\rm p}^{-1}(\hat{\bf f}-
 \o^2\hat{\bf M}_1\hat{\bf L}_{\rm g}^{-1}\langle\hat{\bf L}_{\rm p}^{-1}\hat{\bf
 f}\rangle_2).
\]

\subsection{Periodic lattice and strip with localised
inclusion.}\lb{S43}

In this case the wave equation \er{106} becomes
\[\lb{112}
 \hat{\bf L}\hat{\bf u}=-\o^2(\hat{\bf M}\hat{\bf u}+\hat{\bf
 M}_1\langle\hat{\bf
 u}\rangle_2+\hat{\bf M}_2\langle\hat{\bf u}\rangle)
\]
with $\langle\cdot\rangle\ev\langle\langle\cdot\rangle_1\rangle_2$.
The equation \er{112} is equivalent to
\[\lb{113}
 \hat{\bf L}\hat{\bf u}+\o^2\hat{\bf M}\hat{\bf u}+\o^2\hat{\bf M}_1\langle\hat{\bf
 u}\rangle_2=-\o^2\hat{\bf M}_2\langle\hat{\bf u}\rangle.
\]
Applying \er{110}-\er{111} to \er{113} with $\hat{\bf
f}=-\o^2\hat{\bf M}_2\langle\hat{\bf u}\rangle$ leads to
\[\lb{114}
 \hat{\bf u}={\bf L}_{\rm p}^{-1}
 (\hat{\bf I}-\o^2\hat{\bf M}_1\hat{\bf L}_{\rm g}^{-1}\langle\hat{\bf L}_{\rm p}^{-1}\rangle_2)(-\o^2\hat{\bf M}_2\langle\hat{\bf u}\rangle)
\]
and then
\[\lb{115}
 \langle\hat{\bf u}\rangle=\lt\langle\langle{\bf L}_{\rm
 p}^{-1}\rangle_2
 \lt(\hat{\bf I}-\o^2\hat{\bf M}_1\hat{\bf L}_{\rm g}^{-1}\langle\hat{\bf L}_{\rm p}^{-1}\rangle_2\rt)\rt\rangle_1(-\o^2\hat{\bf M}_2\langle\hat{\bf u}\rangle)
\]
which can be rewritten in the form of
\[\lb{116}
 (\hat{\bf I}+\o^2\langle\hat{\bf L}_{\rm g}^{-1}\langle{\bf L}_{\rm
 p}^{-1}\rangle_2\rangle_1
 \hat{\bf
 M}_2)\langle\hat{\bf u}\rangle=0.
\]
The equation \er{116} gives us the condition of existence of
localised modes \er{013}.

\section{Example}
\setcounter{equation}{0} \lb{S3}
\begin{figure}[h]
\begin{minipage}[h]{0.32\linewidth}
\center{\includegraphics[width=1\linewidth]{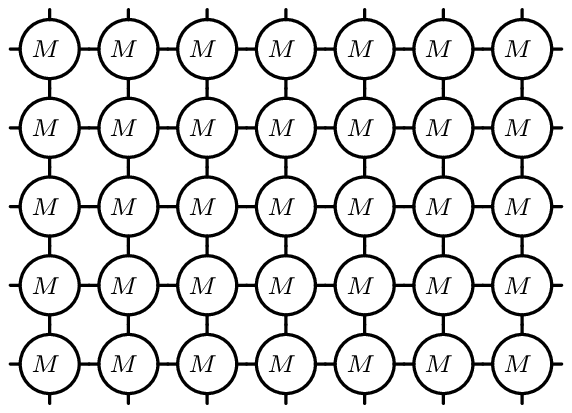}} a) \\
\end{minipage}
\hfill
\begin{minipage}[h]{0.32\linewidth}
\center{\includegraphics[width=1\linewidth]{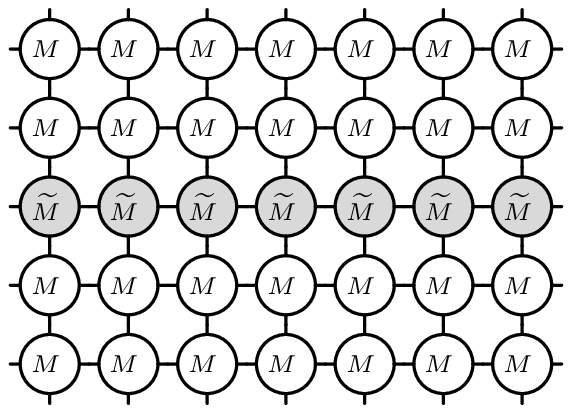}} b) \\
\end{minipage}
\hfill
\begin{minipage}[h]{0.32\linewidth}
\center{\includegraphics[width=1\linewidth]{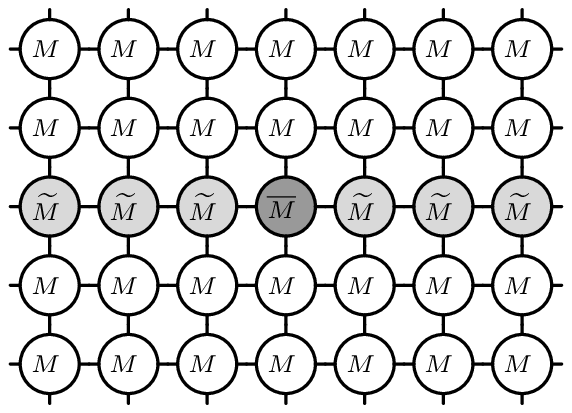}} c) \\
\end{minipage}
\caption{a) Homogeneous lattice with mass $M=1$, b) perturbed by
periodic strip with mass $\wt M=M+M^{(1)}$, c) perturbed by single
mass $\ol{M}=M+M^{(1)}+M^{(2)}$.} \label{fig4}
\end{figure}
\subsection{Uniform lattice.} We start from homogeneous square lattice with mass
$M=1$, see Fig. \ref{fig4}.a). The wave equation for this structure
takes the following form (see \er{003})
\[\lb{300}
 (2\cos k_1+2\cos k_2-4)u=-\o^2u.
\]
The operator $\hat{\bf L}_{\rm p}$ and Floquet branches are (see
\er{009})
\[\lb{301}
 \hat{\bf L}_{\rm p}=\o^2-4+2\cos k_1-2\cos k_2,
\]
\[\lb{302}
 \o_{\rm p}({\bf k})=\sqrt{4-2\cos k_1-2\cos k_2}.
\]
\subsection{Uniform lattice with uniform line.}
Now we add the periodic line with mass $M+M^{(1)}$ (recall that
$M=1$, see Fig. \ref{fig4}.b). In this case
\[\lb{303}
 \langle\hat{\bf L}_{\rm p}^{-1}\rangle_2=\ca
  \frac{-1}{\sqrt{(2\cos k_1-4+\o^2)^2-4}},& {\rm if\ }\o^2<2-2\cos
  k_1,\\
  \frac{1}{\sqrt{(2\cos k_1-4+\o^2)^2-4}},& {\rm if\ }\o^2>6-2\cos
  k_1.
 \ac
\]
Due to \er{010} the guided spectrum $\o_{\rm g}(k_1)$ is determined
by
\[\lb{304}
 (\det\hat{\bf L}_{\rm g}=)\ca
  1-\frac{\o^2M^{(1)}}{\sqrt{(2\cos k_1-4+\o^2)^2-4}}=0,& {\rm if\ }\o^2<2-2\cos
  k_1,\\
  1+\frac{\o^2M^{(1)}}{\sqrt{(2\cos k_1-4+\o^2)^2-4}}=0,& {\rm if\ }\o^2>6-2\cos
  k_1.
 \ac
\]
These equations can be solved directly
\[\lb{305}
 \o_{\rm g}^2(k_1)=\ca
  \frac{2\cos k_1-4-2\sqrt{(M^{(1)})^2(\cos k_1-2)^2-(M^{(1)})^2+1}}{(M^{(1)})^2-1},&{\rm if\ } -1<M^{(1)}<0,\\
  \frac{2\cos k_1-4+2\sqrt{(M^{(1)})^2(\cos k_1-2)^2-(M^{(1)})^2+1}}{(M^{(1)})^2-1},&{\rm if\ } M^{(1)}>0.
 \ac
\]
The projection of the propagative spectrum $\o_{\rm p}({\bf k})$
\er{302} on the plane $(\o,k_1)$ (see \er{012}) is
\[\lb{306}
 I_{\rm p}(k_1)=\lt[\sqrt{2-2\cos k_1},\sqrt{6-2\cos k_1}\rt].
\]
\begin{figure}[h]
\begin{minipage}[h]{0.241\linewidth}
\center{\includegraphics[width=1\linewidth]{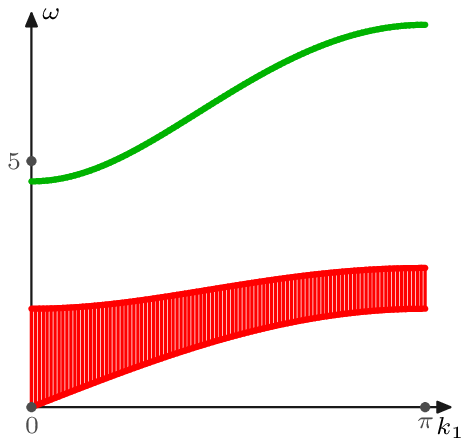}} \small{\small a)} $\scriptstyle M^{(1)}=-0.9$ \\
\end{minipage}
\hfill
\begin{minipage}[h]{0.241\linewidth}
\center{\includegraphics[width=1\linewidth]{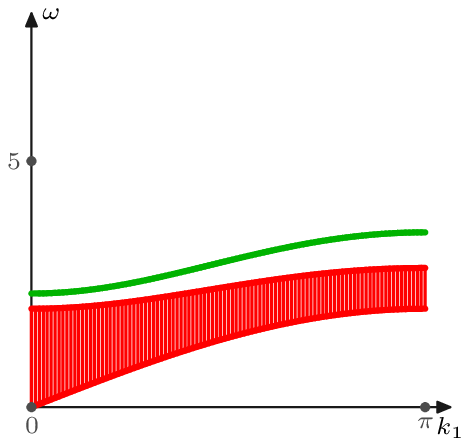}} \small{\small b)} $\scriptstyle M^{(1)}=-0.5$ \\
\end{minipage}
\hfill
\begin{minipage}[h]{0.241\linewidth}
\center{\includegraphics[width=1\linewidth]{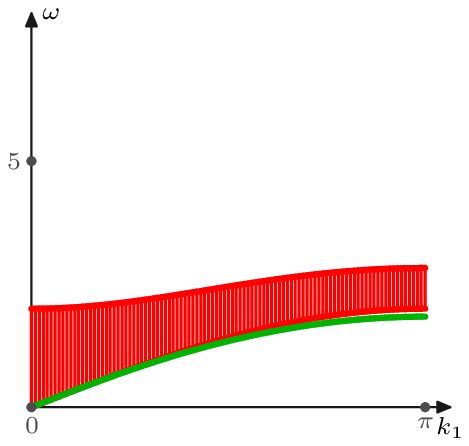}} \small{\small c)} $\scriptstyle M^{(1)}=0.5$ \\
\end{minipage}
\hfill
\begin{minipage}[h]{0.241\linewidth}
\center{\includegraphics[width=1\linewidth]{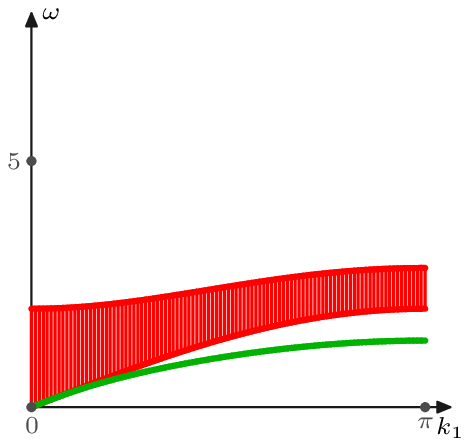}} \small{\small d)} $\scriptstyle M^{(1)}=2$ \\
\end{minipage}
\caption{Projection of the Floquet spectrum $I_{\rm p}(k_1)$
\er{306} on the plane $(\o,k_1)$ (red area) and the guided spectrum
(green line) $\o_{\rm g}(k_1)$ \er{305} for homogeneous square
lattice (mass $M=1$) with perturbed line (mass $\wt M=M+M^{(1)}$,
see Fig. \ref{fig4}.b).} \label{fig1}
\end{figure}

\subsection{Uniform lattice and line with localised
inclusion.} Now we add one perturbed mass $\ol{M}=\wt M+M^{(2)}$.
The condition on $\o_{\rm loc}$ (see \er{013}) takes the following
form
\[\lb{307}
 D_{\rm
 loc}(\o)\ev1+\frac{\o^2M^{(2)}}{2\pi}\int_{-\pi}^{\pi}\frac{dk_1}{\o^2M^{(1)}+\sqrt{(2\cos
 k_1-4+\o^2)^2-4}}=0.
\]
The projection of the propagative and guided spectrum on the axis
$\o$ (see \er{015}) is
\begin{eqnarray}\lb{308}
 I_{\rm p}&=&[0,2\sqrt{2}],\\
 I_{\rm g}&=&\ca \lt[\frac2{\sqrt{1-(M^{(1)})^2}},\frac{\sqrt{6+2\sqrt{8(M^{(1)})^2+1}}}{\sqrt{1-(M^{(1)})^2}}\rt],&{\rm if\ } -1<M^{(1)}<0,\\
                            \lt[0,\sqrt{\frac{-6+2\sqrt{8(M^{(1)})^2+1}}{(M^{(1)})^2-1}}\rt], &{\rm if\ }
                            M^{(1)}>0.
                        \ac
\end{eqnarray}

\begin{figure}[h]
\begin{minipage}[h]{0.241\linewidth}
\center{\includegraphics[width=1\linewidth]{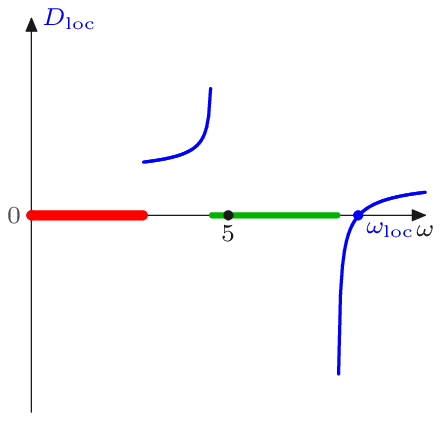}} \small{\small a)} $\scriptstyle M^{(1)}=-0.9,\ M^{(2)}=-0.03$ \\
\end{minipage}
\hfill
\begin{minipage}[h]{0.241\linewidth}
\center{\includegraphics[width=1\linewidth]{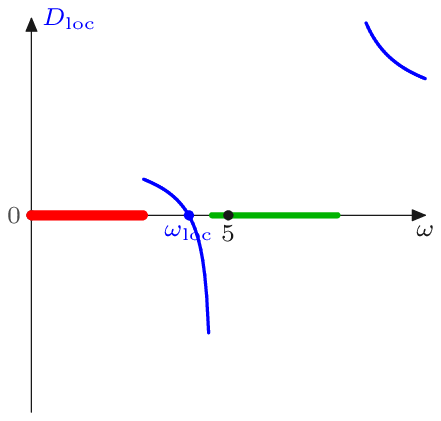}} \small{\small b)} $\scriptstyle M^{(1)}=-0.9,\ M^{(2)}=0.1$ \\
\end{minipage}
\hfill
\begin{minipage}[h]{0.241\linewidth}
\center{\includegraphics[width=1\linewidth]{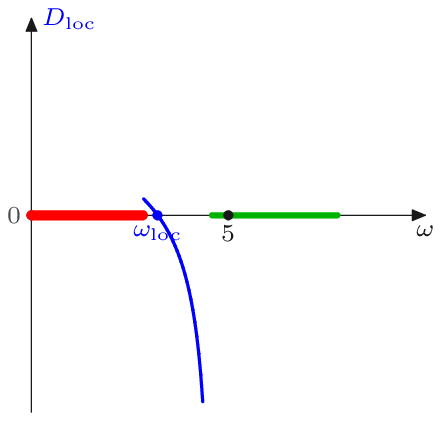}} \small{\small c)} $\scriptstyle M^{(1)}=-0.9,\ M^{(2)}=0.25$ \\
\end{minipage}
\hfill
\begin{minipage}[h]{0.241\linewidth}
\center{\includegraphics[width=1\linewidth]{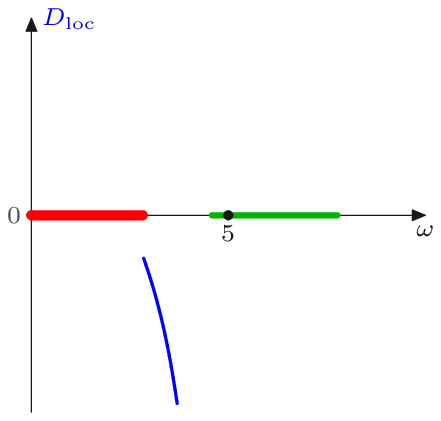}} \small{\small d)} $\scriptstyle M^{(1)}=-0.9,\ M^{(2)}=0.7$ \\
\end{minipage}
\vfil
\begin{minipage}[h]{0.241\linewidth}
\center{\includegraphics[width=1\linewidth]{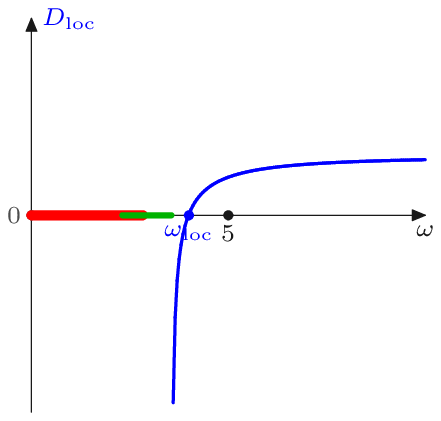}} \small{\small e)} $\scriptstyle M^{(1)}=-0.5,\ M^{(2)}=-0.2$ \\
\end{minipage}
\hfill
\begin{minipage}[h]{0.241\linewidth}
\center{\includegraphics[width=1\linewidth]{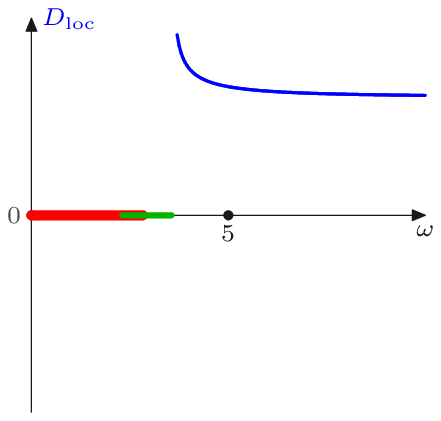}} \small{\small f)} $\scriptstyle M^{(1)}=-0.5,\ M^{(2)}=0.1$ \\
\end{minipage}
\hfill
\begin{minipage}[h]{0.241\linewidth}
\center{\includegraphics[width=1\linewidth]{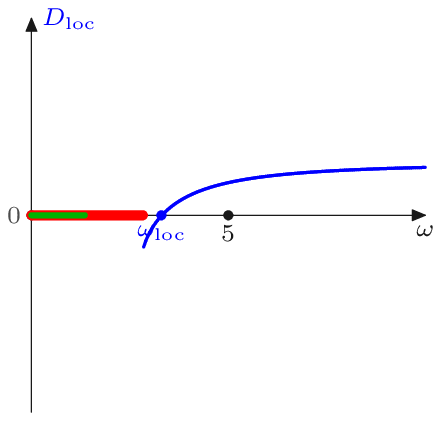}} \small{\small g)} $\scriptstyle M^{(1)}=2,\ M^{(2)}=-2.6$ \\
\end{minipage}
\hfill
\begin{minipage}[h]{0.241\linewidth}
\center{\includegraphics[width=1\linewidth]{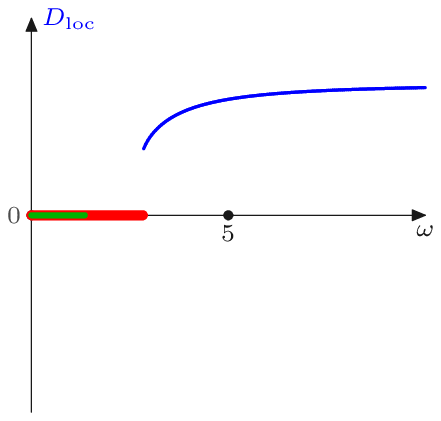}} \small{\small h)} $\scriptstyle M^{(1)}=2,\ M^{(2)}=-2$ \\
\end{minipage}
\caption{Function $D_{\rm loc}(\o)$ \er{307} (blue curve) and its
zeroes $\o_{\rm loc}$ (localised spectrum) with the projection of
propagative $I_{\rm p}$ \er{308} (red bold line) and guided $I_{\rm
g}$ \er{308} (green thin line) spectra on the axe $\o$.}
\label{fig2}
\end{figure}

\subsection{Existence of the localised waves (states).} The equation \er{307}
can be rewritten in the form
\[\lb{309}
 D_{\rm loc}(\o)=1+M^{(2)}D_1(\o)=0
\]
with
\[\lb{310}
 D_1(\o)\ev\frac{1}{2\pi}\int_{-\pi}^{\pi}\frac{\o^2dk_1}{\o^2M^{(1)}+\sqrt{(2\cos
 k_1-4+\o^2)^2-4}}.
\]
The integrand function in \er{310} is strictly decreasing in
$\o\in\R_+\sm(I_{\rm p}\cup I_{\rm g})$ for any $k_1$. Then
$D_1(\o)$ is strictly decreasing function in $\o\in\R_+\sm(I_{\rm
p}\cup I_{\rm g})$. So knowing the values of $D_1(\o)$ at the edges
of $I_{\rm p}$ and/or $I_{\rm g}$ and using \er{309} with
monotonicity of $D_1(\o)$ we can predict the existence of localised
modes in the corresponding gap. It is convenient to consider
different cases (recall that $1+M^{(1)}>0$ and $1+M^{(1)}+M^{(2)}>0$
because these are the masses of inclusions):

\subsubsection{The case $M^{(1)}\in(-1,-1/\sqrt{2})$.} Due to
\er{308} we have two disjoint gaps $\cG_1$ and $\cG_2$ in the
propagative and guided spectrum, i.e.
\[\lb{311}
 \R_+\sm(I_{\rm p}\cup I_{\rm g})=\cG_1\cup\cG_2,\ \ {\rm where}
\]
\[\lb{312a}
 \cG_1=\lt(2\sqrt{2},\frac2{\sqrt{1-(M^{(1)})^2}}\rt),\ \
 \cG_2=\lt(\frac{\sqrt{6+2\sqrt{8(M^{(1)})^2+1}}}{\sqrt{1-(M^{(1)})^2}},+\iy\rt).
\]
The values of $D_1$ at the edges of gaps are
\begin{eqnarray}
D_1(2\sqrt{2})&=&\frac{4}{\pi}\int_{0}^{\pi}\frac{dk_1}{4M^{(1)}+\sqrt{(\cos
k_1+2)^2-1}}<0,\lb{312}\\
D_1\lt(\frac2{\sqrt{1-(M^{(1)})^2}}\rt)&=&-\iy,\lb{313}\ \
D_1\lt(\frac{\sqrt{6+2\sqrt{8(M^{(1)})^2+1}}}{\sqrt{1-(M^{(1)})^2}}\rt)=+\iy,\lb{314}\\
D_1(+\iy)&=&\frac1{1+M^{(1)}}>0.\lb{315}
\end{eqnarray}
Using \er{312}-\er{315}, \er{309} with monotonicity of $D_1$ we
conclude that:

{\it a) If $M^{(2)}<0$ then there are no localised states in $\cG_1$
and there is only one localised state in $\cG_2$.

b) If $M^{(2)}=0$ then there are no localised states in $\cG_1$ and
in $\cG_2$.

c) If $M^{(2)}\in(0,-1/D_1(2\sqrt{2}))$ then there is only one
localised state in $\cG_1$ and no localised states in $\cG_2$.

d) If $M^{(2)}\ge -1/D_1(2\sqrt{2})$ then there are no lacalised
states in $\cG_1$ and $\cG_2$.}

\subsubsection{The case $M^{(1)}\in[-1/\sqrt{2},0]$.} Due to
\er{308} we have only one gap $\cG_2$ in the propagative and guided
spectrum, i.e.
\[\lb{316}
 \R_+\sm(I_{\rm p}\cup I_{\rm g})=\cG_2=
 \lt(\frac{\sqrt{6+2\sqrt{8(M^{(1)})^2+1}}}{\sqrt{1-(M^{(1)})^2}},+\iy\rt).
\]
Applying the same arguments as in the previous case we deduce that:

{\it a) If $M^{(2)}<0$ then there is only one localised state in
$\cG_2$.

b) If $M^{(2)}\ge0$ then there are no localised states in $\cG_2$. }

\subsubsection{The case $M^{(1)}>0$.} Due to
\er{308} we have only one gap $\cG$ in the propagative and guided
spectrum, i.e.
\[\lb{317}
 \R_+\sm(I_{\rm p}\cup I_{\rm g})=\cG=
 (2\sqrt{2},+\iy).
\]
Applying the same arguments as in the previous cases we deduce that:

{\it a) If $M^{(2)}<-1/D_1(2\sqrt{2})$ then there is only one
localised state in $\cG$.

b) If $M^{(2)}\ge-1/D_1(2\sqrt{2})$ then there are no localised
states in $\cG$.}

Note that in this case $D_1(2\sqrt{2})>0$ (compare with \er{312}).

\subsubsection{Summation.} Now we summarize above statements.

\begin{figure}[h]
\begin{minipage}[h]{0.49\linewidth}
\center{\includegraphics[width=1\linewidth]{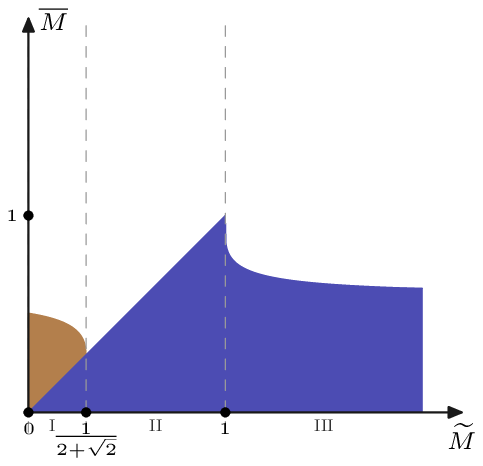}}
\end{minipage}
\hfill
\begin{minipage}[h]{0.49\linewidth}
1) For the strip "I" there are two spectral gaps, first gap is
located between propagative and guided spectra, the second is above
the guided spectrum (Fig. \ref{fig1}.a). For the brown area the
localised mode is in the first gap (Figs. \ref{fig2}.b,c), for the
purple one is in the second (Fig. \ref{fig2}.a).

2) For the strip "II" there is one spectral gap above the guided
spectrum. Guided and propagative spectra intersect (Fig.
\ref{fig1}.b). For the purple area there is one localised mode in
the gap (Fig. \ref{fig2}.e).

3) For the quadrant "III" there is one gap above the propagative
spectrum. Guided spectrum lies inside the propagative (Figs.
\ref{fig1}.c,d). For the purple area there is one localised mode in
the gap (Fig. \ref{fig2}.g).

For the white area we have no localised modes in the spectral gaps
(Figs. \ref{fig2}.d,f,h).

\end{minipage}
\caption{The regions of emergence of localised modes for different
masses $\wt M$, $\ol{M}$ in the lattice Fig. \ref{fig4}.c). The
boundary of these regions consist of curves
$\ol{M}=\wt{M}-\frac1{D_1(2\sqrt{2})}$ (see \er{312} for $D_1$) and
$\ol{M}=\wt{M}$. The maximal mass $\ol{M}$ for which the localised
mode exists tends to $1$ (with $\wt M\to1$). After increasing of
mass $\wt M\ge1$ we see at the beginning the rapid fall of mass
$\ol{M}$, which reach the limit $\frac34-\frac1{2\pi}$ for $\wt
M\to+\iy$.} \label{fig3}
\end{figure}

\section{Conclusion.}

The dispersion equations for the propagative, guided and localised
waves in the discrete periodic lattice with the strip and with
localised inclusions are obtained. For the uniform lattice with the
line and with one single inclusion the existence of localised modes
and its position is analysed.

{\bf Acknowledgement.} The author is grateful to A. Shuvalov for
useful discussions.


\end{document}